# Keep it green, simple and socially fair: a choice experiment on prosumers' preferences for peer-to-peer electricity trading in the Netherlands

Elena Georgarakis[a], Thomas Bauwens[a], Anne-Marie Pronk[b], Tarek AlSkaif[c]

[a] Copernicus Institute of Sustainable Development, Utrecht University, 3584 CS Utrecht, the Netherlands

[b] EnergyCoin Foundation, Arnhem, the Netherlands

[c] Information Technology Group, Wageningen University and Research, 6706 KN Wageningen, the Netherlands

Corresponding address: Elena Georgarakis, Faculty of Geosciences, Utrecht University, Princetonlaan 8A, 3584 CS Utrecht, the Netherlands. Email: e.z.georgarakis@uu.nl

## Abstract

While the potential for peer-to-peer electricity trading, where households trade surplus electricity with peers in a local energy market, is rapidly growing, the drivers of participation in this trading scheme have been understudied so far. In particular, there is a dearth of research on the role of non-monetary incentives for trading surplus electricity, despite their potentially important role. This paper presents the first discrete choice experiment conducted with prosumers (i.e. proactive households actively managing their electricity production and consumption) in the Netherlands. Electricity trading preferences are analyzed regarding economic, environmental, social and technological parameters, based on survey data (N = 74). The dimensions most valued by prosumers are the environmental and, to a lesser extent, economic dimensions, highlighting the key motivating roles of environmental factors. Furthermore, a majority of prosumers stated they would provide surplus electricity for free or for non-monetary compensations, especially to energy-poor households. These observed trends were more pronounced among members of energy cooperatives. This suggests that peer-to-peer energy trading can advance a socially just energy transition. Regarding policy recommendations, these findings point to the need for communicating environmental and economic benefits when marketing

P2P electricity trading platforms and for technical designs enabling effortless and customizable transactions.

# 1. Introduction

By signing the Paris Agreement, 196 nations committed to pursuing efforts that limit the increase of the global average temperature to 1.5°C above pre-industrial levels, reducing the risks and impacts of climate change (United Nations, 2016). An integral part of the Paris Agreement is a transition to renewable energies and, therewith, the promotion of distributed energy resources (DER), defined as electric power generating resources that are directly connected to a medium- or low-voltage distribution network (Akorede et al., 2010). The increasing deployment of DERs comes with opportunities and challenges. Although they can increase efficiency and reduce $CO_2$ emissions when appropriately managed (Akorede et al., 2010), this management requires advanced mechanisms. A centralized electricity grid will face issues regarding the balance of supply and demand when more DERs from variable energy sources are installed (Rommel and Sagebiel, 2017). Additionally, in some countries such as the Netherlands[1], feeding surplus energy into centralized grids is becoming less economically attractive for prosumers (i.e. proactive households managing their electricity production and consumption) due to changes in support mechanisms.

In light of this evolution, decentralized smart grids, relying on technologies such as smart meters and platforms enabling the generation, consumption and trading of electricity, are receiving increasing attention, as they provide the opportunity to form local electricity markets (LEM) (Parag and Sovacool, 2016). LEM, in which prosumers can trade surplus electricity directly with a community or peers in peer-to-peer (P2P) trading schemes, represent an alternative to the dependence on centralized grids and their hierarchical structure (Crespo-Vazquez et al., 2020). P2P trading schemes present many potential benefits on the individual and societal level. When surplus electricity is traded P2P, individual prosumers can optimize the utilization of the DER in terms of financial returns and the balance of supply and demand, based on actual and predicted energy prices, as well as generation and loads (van Leeuwen et al., 2020). This can reduce demand costs for the energy consumer, because less electricity from the central grid is required (AlSkaif et al., 2017; Lüth et al., 2018), while generating environmental benefits for society at large in the form of reduced grid electricity from fossil resources (depending on the prevailing energy mix) and reduced energy losses linked to long transmission distances (Jogunola et al., 2017).

---

[1] In the Netherlands, the current financially attractive net-metering system will be replaced by *terugleversubsidie,* a return subsidy comparable to a feed-in tariff (FiT) which will effectively decrease the financial return for prosumers (Rijksoverheid, 2019).

Studying the preferences for P2P electricity trading is relevant to maximize social welfare, because prosumers typically have heterogeneous preferences and are willing to pay different prices to exchange energy, depending on factors such as generation technology, location in the network and the owner's reputation (Morstyn and McCulloch, 2019). Moreover, providing additional reasons, beyond purely financial drivers, for participating in P2P electricity trading may contribute to closing the so-called intention-behavior gap (i.e. the discrepancy between an individual's intentions and their actual behavior; Claudy et al., 2014). Yet, while many studies have looked at the technical solutions required to successfully implement a P2P trading platform (e.g. Morstyn and McCulloch, 2019; van Leeuwen et al., 2020 ), less research has focused on the preferences for P2P electricity trading (Hackbarth and Löbbe, 2020). Furthermore, the few existing studies on drivers and barriers for households to engage in P2P electricity trading are based on survey data of individual households - regardless of whether they are prosumers or regular energy users - covering only a few countries, including Germany, Switzerland and Australia (Ecker et al., 2018; Hackbarth and Löbbe, 2020; Hahnel et al., 2020; Mengelkamp et al., 2019; Mengelkamp et al., 2018; Reuter and Loock, 2017; Wilkinson et al., 2020). In addition, while some studies consider certain social factors, such as community identity and social equity, as potential drivers for the participation in P2P electricity markets (Hackbarth and Löbbe, 2020; Mengelkamp et al., 2018; Wilkinson et al., 2020), the social dimension remains relatively understudied to date.

This paper tackles these research gaps by considering the environmental, economic, technological, and social factors that drive the acceptance and, thereby, the success of P2P electricity trading systems. It extends the existing scientific literature on individuals' preferences for P2P electricity trading in several ways. First, it solely focuses on prosumers. This is expected to yield more valid results, because respondents are likely to have more insight into the current and alternative processes of electricity trading than regular energy users. Furthermore, studying prosumers' preferences is especially relevant because prosumers both consume (buy) and also provide (sell) electricity. This is necessary to generate a local supply of electricity and create a P2P energy market. Second, it assesses the trading preferences of prosumers with diverse characteristics and attitudes. In particular, it compares the preferences of individual prosumers to those of energy cooperatives members (an emblematic form of collective prosumerism; Campos et al., 2020) in order to shed light on the potential role of these organizations in diffusing P2P electricity trading. Third, it surveys prosumers located in the Netherlands, providing insights from a geographical setting that has not been studied before. Fourth, based on insights from anthropological research (Singh et al., 2018), the current study offers a more nuanced depiction of the social drivers of the participation in P2P trading by considering different types of returns (including in-cash, in-kind and intangible, non-monetary compensations) as well as different types of social relationships.

To achieve this, a questionnaire-based survey including a discrete choice experiment (DCE) with prosumers in the Netherlands was conducted. The survey was used to investigate variations in trading preferences of prosumer subgroups and the willingness to exchange surplus electricity for free or non-monetary compensations. The final sample consisted of 74 prosumers, 45% of whom were members of an energy cooperative.

The rest of the article is structured as follows. Section 2 provides an overview of the relevant literature on prosumers, energy communities and cooperatives, and P2P electricity trading. In Section 3, the design process of the survey and the data collection process are described. Section 4 presents and discusses the results obtained through the analysis of the questionnaire-based survey and the DCE. Section 5 gives a conclusion of the main findings and provides policy recommendations.

## 2. Conceptual Framework

2.1. Individual and collective prosumers

Among prosumers willing to produce energy to provide for their own energy needs and to participate in energy markets, one can distinguish between individual prosumers, who install their own DER system (e.g. solar panels, micro-wind turbines), and collective prosumers, which refer to some form of organization (i.e. companies, municipalities, energy cooperatives, etc.) gathering multiple individuals who jointly co-own and manage DER (Inês et al., 2020). Energy cooperatives, which are organizations characterized by democratic decision-making and fair distribution of the economic surplus (Bauwens, 2016), have been recognized as the most common form of collective prosumer initiatives in Europe and their number is increasing swiftly in countries like Germany, the UK, Belgium and the Netherlands (Bauwens et al., 2016; Campos et al., 2020). Furthermore, energy cooperatives were identified as facilitating actors in the transition to low-carbon energy systems (Seyfang et al., 2013) and may therefore also facilitate and drive the establishment and adoption of P2P electricity trading.

Both individual and collective prosumers are potential participants in P2P electricity trading. Previous research projects which investigated preferences of prosumers regarding the way they want to trade and share self-produced electricity has highlighted a multiplicity of drivers and barriers. This paper follows the lines of reasoning adopted by several previous studies (e.g. Hahnel et al., 2020; Mengelkamp et al., 2019; Mengelkamp et al., 2018) by examining these drivers and barriers along the economic, environmental, social and technological features of P2P electricity trading. These dimensions are reviewed in the following section.

2.2. Drivers and barriers for participation in P2P electricity trading

**2.1.1. Economic**

Individual economic benefits, for example in the form of cost savings from their self-produced electricity, have been identified as important motivations for individuals to install DER and thereby becoming a prosumer (Palm, 2018). The introduction of subsidies, such as feed-in tariffs, has also encouraged the adoption of microgeneration, for example in the UK (Balcombe et al., 2014, 2013) and in Germany (Schaffer and Brun, 2015). In this perspective, a good knowledge about the involved costs and the existing subsidies showed significant positive impact on adoption of photovoltaic systems (Vasseur and Kemp, 2015), while perceived uncertainty around regulations and difficulties in the process of feeding-in and selling surplus electricity were identified as barriers for becoming a prosumer (Palm, 2017). Regarding collective economic benefits, economic incentives, including a high return on investment and a lower electricity price, were also identified as key drivers to join energy cooperatives (Bauwens, 2019). By contrast, economic barriers include high investment costs, long pay-off time and lack of subsidies (Balcombe et al., 2014, 2013; Enlund and Eriksson, 2016; Palm and Tengvard, 2017).

Some studies found economic aspects to be the most important factor in decision-making processes regarding P2P electricity trading (Hackbarth and Löbbe, 2020; Mengelkamp et al., 2019). Economic benefits of participation in LEM with P2P trading are achieved by an optimal utilization of renewable DER and the related optimization of electricity procurement costs and returns, as losses can be minimized and demand may be reduced (Jogunola et al., 2017). The economic aspects of P2P electricity are also reflected by the required time to manage transactions, as time can be seen as a scarce resource that households treat similarly to money (Heckman, 2015). In contrast to automatically feeding surplus electricity into the central electricity grid, participation in LEM requires regular interaction with an information system and therefore entails additional time and effort (Mengelkamp et al., 2018). Thus, easing the implementation and comfortable use of such systems through well-designed algorithms can result in increased willingness to participate in P2P electricity trading (Hackbarth and Löbbe, 2020).

### 2.1.2. Environmental

Environmental concerns, such as the desire to live a more sustainable lifestyle, have consistently been identified by previous studies as a major driver for individuals that decide to install renewable DER (Palm, 2018; Wittenberg and Matthies, 2016). Similarly, several studies looking at the motivations to join collective prosumer initiatives find that environmental concerns play a major role (Bauwens, 2019). For instance, looking at individual motivations to join renewable energy cooperatives in Flanders, Bauwens (2017) showed that the support for the production of renewable energy was a more important motivation for joining such initiatives than the return on investment or the electricity price. Relying on a survey among members of German community-based renewable energy initiatives, Radtke (2014) showed that participants' involvement was primarily driven by environmental motivations rather than financial return. A number of studies also confirmed that environmental benefits are the most influential factor for the participation in P2P electricity markets (Hackbarth and Löbbe, 2020; Mengelkamp et al., 2018; Reuter and Loock, 2017; Wilkinson et al., 2020).

**2.1.3. Social**

Social factors play an essential role in motivating individuals to become prosumers. In particular, various studies have shown the importance of peer effects and social influences in the adoption of renewable DER such as solar PV (Axsen et al., 2013; Bollinger and Gillingham, 2012; Palm, 2017). Peer effects are the influence of a person's peers (e.g. neighbors, friends, relatives or colleagues) on their behavior and mainly occur through existing and rather close social relationships. They can facilitate the diffusion of DER by reducing barriers to adoption and by enhancing the level of trust in a technology (Palm, 2017).

Research focusing on the drivers to join energy cooperatives also emphasizes the role of social factors, in particular, social identity to the group. Social identity can be defined as "that part of an individual's self-concept which derives from his knowledge of his membership of a social group (or groups) together with the value and emotional significance attached to that membership" (Tajfel, 1978). Bauwens (2016) shows that energy cooperative members who identify more strongly with their cooperative tend to invest larger amounts of money and to participate more frequently in annual general assemblies. Regarding participation in LEM and P2P trading, Mengelkamp et al. (2018) also found a positive influence from individuals' sense of community identity. Similarly, Wilkinson et al. (2020), in their study of a P2P market trial in Australia, found that many participants joined the trial on the basis of supporting the local community and broader social equity issues. This suggests that trade partners' identity as well as the way the market is designed matters to prosumers' decision-making processes when they trade surplus electricity. The ability to decide whom to share their electricity with is therefore an asset and a core feature of electricity trading platforms designed for P2P electricity exchange.

The preference for the social connection with trading partners may also be reflected in the participant's preferred return. In a field research carried out in rural India, Singh et al. (2018) found that preferred returns for energy provided to peers vary depending on the prosumer's personal relationship with their peer or community. The authors found that the closer the connection of energy providers was with the consumer, the more likely they accepted returns different from in-cash payments: *in-kind returns* of a non-cash form but still of monetary value, and *intangible returns* of a non-monetary form, which are unmeasurable and unquantifiable social gestures. Accordingly, P2P energy trading can be an opportunity to provide electricity to everyone in a more equal and just manner (Giotitsas et al., 2015; Ruotsalainen et al., 2017). The concept can be used to encourage sharing resources for the benefit of individuals, communities and society at large, leading to more energy efficiency and a democratization of energy (Parag and Sovacool, 2016).

**2.1.4. Technological**

Further motives for becoming a prosumer are independence from energy companies and curiosity about the technology, while inexperience with underlying technologies and their installation is perceived as a barrier (Palm, 2018). A technological interest also influences the participation in LEM positively (Mengelkamp et al., 2018; Reuter and Loock, 2017; Wilkinson et al., 2020). P2P trading can increase prosumers' self-sufficiency allowing for a certain level of autonomy from private or state-owned energy suppliers and grid operators (Morstyn et al., 2018). At the same time, distributed ledger technologies behind the electricity trading platform ensure that data is kept private and protected (AlSkaif et al., 2021; Buth et al., 2019).

When appropriately managed, local and smart electricity grids have the potential to improve grid efficiency and to utilize more of the energy that is generated by a renewable DER, thereby reducing energy losses (Morstyn et al., 2018) and the required imports from a central grid (van Leeuwen et al., 2020). P2P trading can help to increase the overall efficiency of the energy system due to locally optimized management of supply and demand. Recent studies show that considering energy efficiency improvements in P2P trading can result in a better match between demand and local energy supply, which can be considered as a proxy for reducing losses in the network (AlSkaif et al., 2021). Furthermore, by optimizing the utilization of DER to match the flexible demand (e.g. charging electric vehicles), peak load can be reduced (Brinkel et al., 2020).

The four dimensions of drivers and barriers identified in literature are expected to have a varying influence on prosumers' individual preferences for trading electricity. In particular, we expect that environmental and economic factors play the most important role, in line with previous studies. The next section presents the methods used to analyze this influence, to identify differences in trading preferences in prosumers with diverse characteristics and attitudes, (e.g. regarding the membership in an energy cooperative) and to examine the potential of non-monetary compensation for trading surplus electricity.

## 3. Methodology and Data

### 3.1. Context

The survey underlying this research was conducted with prosumers in the Netherlands. Three trends in the Netherlands provide a strong case for alternative electricity trading schemes like P2P trading and, therefore, motivated the choice of this country as the empirical setting for this study. First, the Dutch government made commitments to increase the share of renewable energy to at least 27% in 2030 (Ministry of Economic Affairs and Climate Policy, 2019). Second, in 2023, the current financially attractive net-metering system will be replaced by *terugleversubsidie,* a return subsidy comparable to a feed-in tariff (FiT), which will effectively decrease the financial return for prosumers (Rijksoverheid, 2019). Third, the number of energy cooperatives is increasing rapidly. In 2019, the number of energy cooperatives in the Netherlands increased by 20% compared to the previous year – accumulating to 582 energy cooperatives with an estimate of 85,000 members (HIER opgewekt and RVO, 2019). Some regulations have contributed to this development, notably the postal-code-area regulation, which enables local energy cooperatives to supply their members with electricity (ECoop, 2020). The regulation stipulates that members of an energy cooperative must share the same postal code in order to exchange electricity among themselves, and to apply to the net metering law and be eligible for tax advantages (Campos et al., 2020; Kooij et al., 2018). This exemption allows prosumers to participate in retail electricity markets, for which they usually need to have a status as a supplier (Campos et al., 2020).

3.2. Design of the discrete choice experiment

This research made use of a DCE, which is a widely applied method for analyzing complex decision-making processes (Mangham et al., 2009). The method uses an attribute-based survey to measure individual preferences. It assumes that participants strive to maximize their utility through rational decision-making when selecting a bundle of attributes which satisfies their needs. The method strives to simulate real-life decision-making processes in order to calculate preference scores which respondents assign to the selected attributes that define the product or service (Lüthi and Prässler, 2011).

For the DCE, several key attributes were selected to describe the different alternatives of electricity trading systems. Attributes should reflect the characteristics of the investigated decision that are expected to affect respondents' choices the most. At the same time attributes should be policy relevant (Mangham et al., 2009). The selected attributes (see Table 1) are all deemed appropriate to investigate the preferences of prosumers regarding electricity exchange. They cover economic, environmental, social and technological aspects that are related to renewable DER and electricity trading in order to incorporate the multiple dimensions that prosumers may consider when making a decision regarding the use of surplus electricity. A participant repeatedly choosing beneficial attribute levels of one aspect over others can be interpreted as the participant preferring that aspect. Each attribute has several levels, which define the attributes' possible values. Overall, six attributes were chosen, and three levels were assigned to each attribute. The levels were chosen such that one describes a trading process with the traditional grid, one means to reflect a P2P trading system, and one describes an intermediate state reflecting indifference. The first selection of attributes was made from reviewing relevant literature. To identify the attributes and attribute levels most relevant for answering the research question, experts and prosumers were interviewed.

Table 1. Final selection of attributes from the relevant literature on P2P electricity trading, the researcher's and the project partners' prior knowledge and experience on the topic, and through interviews with prosumers and experts.

| *Attribute* | *Category* | *Levels* | | |
|---|---|---|---|---|
| 1  $CO_2$ emissions | Environmental | Low | Medium | High |
| 2  Social connection with electricity trading partner | Social | None (anonymous) | Direct (neighbor) | Close (friends and family) |
| 3  Selling price | Economic | 10 €ct/kWh | 15 €ct/kWh | 20 €ct/kWh |
| 4  Additional effort (time) | Economic | 0 h/month | 2 h/month | 4 h/month |
| 5  Improved efficiency | Technological | 5% | 15% | 30% |
| 6  Self-sufficiency | Technological | Low | Medium | High |

In so-called choice sets, participants are presented with three alternative trading scenarios – each consisting of all six attributes with varying attribute levels. Most DCEs do not contain more than sixteen choice sets to remain below the "boredom threshold", which describes the point at which questionnaire participants become fatigued from being asked too many questions (Hanson et al., 2005). As the survey in this study does include several questions besides the DCE, the decision was made to include only twelve choice tasks. Of the twelve choice sets, nine were designed to later analyze participants' preferences and three were designed as so-called holdout tasks. The nine choice sets were designed making use of the Balanced Overlap function in the Lighthouse Studio software tool (Sawtooth Software, 2020)[2]. The remaining three choice sets are so-called holdout tasks and serve as a validity measure. These choice sets are fixed and thereby every participant received them in the same position within the choice experiment. Considering the number of alternative options in a choice task, this DCE used three alternatives which corresponded with the amount of levels in each attribute. Three alternatives generate sufficient data for analyses, while also being processed well by participants (Orme, 2010).

3.3 Questionnaire structure

The questionnaire consisted of four sections as illustrated in Figure 1. The first section of the survey was intended to collect insights about participants' characteristics through several closed-ended questions that they could answer by choosing one or several options from a predefined set of answers (see Appendix Table A.1). The questions covered the reasons why they decided to install a renewable energy system, the reasons why they chose their current energy supplier, whether they were members of an energy cooperative and, if so, the reasons why they became a member and, finally, their willingness to give surplus electricity for free or for indirect financial compensation. The second section contained the DCE as described in the previous section (see Appendix Table A.2).[3]

    The third section intended to measure the participants' level of environmental concern, their sense of identity with their local community, and their affinity towards technology (see Table 2, Appendix Table A.3). Each of these three constructs were assessed with two statement questions,

---

[2] The Balanced Overlap function generates a randomised design, which results in a large number of unique choice sets so that each participant received a unique DCE. This has the advantage of generating significantly more data than in a fixed design where every participant would receive the same choice sets.

[3] A video showing how the DCE is answered by an exemplary respondent is provided with the supplementary material of this article.

which were to be answered on a 5-point Likert scale which ranged from 1 (strongly disagree) to 5 (strongly agree). The internal consistency of the two statements measuring each construct was assessed with Cronbach's Alpha. With values above 0.6 for all three constructs, they are internally consistent, i.e. the two chosen statements are reliably measuring the same concept (Peterson, 1994). The fourth and last section of the survey contained questions regarding the respondent's socio-economic characteristics (see Appendix Table A.4) asking participants about their socio-economic background, such as their age, gender, and education. Furthermore, they were asked in which setting, i.e. large city, medium city, small city or village, their household is located, how many people live in their household, and their household's average yearly net income.

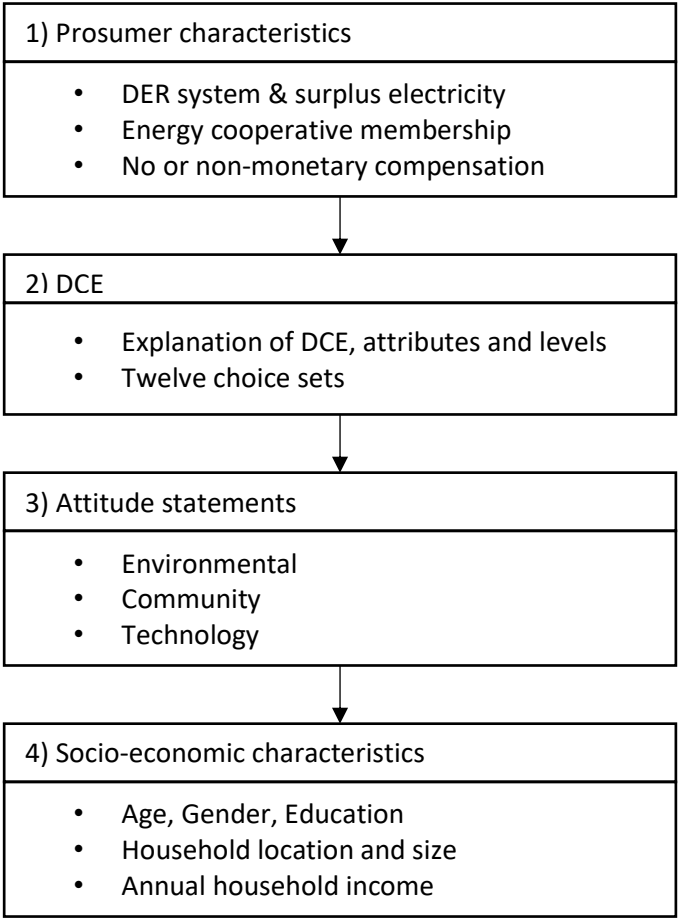

Figure 1. Schematic representation of the questionnaire structure.

Table 2. Statements regarding attitude towards the environment, their local community, and technology.

| Construct | Cronbach's alpha | Statement | Source |
|---|---|---|---|
| *Environmental concern* | 0.65 | I am concerned about human behavior and its impact on the climate and the environment. | Hackbarth and Löbbe (2020) after Kuckartz (2000) |
| | | I always pay attention to ecological criteria when buying products and services. | |
| *Community identity* | 0.82 | I feel a strong identification with my local community. | Mengelkamp et al. (2019) |
| | | There are many people in my local community whom I think of as friends. | Kalkbrenner and Roosen (2016) |
| *Affinity to technology* | 0.83 | Learning how to use a technological device is easy for me. | Karrer et al. (2009) |
| | | I enjoy exploring new technologies. | |

3.4 Data collection and sample

The link to the online survey was distributed through the network of the project partners via email and social media to reach as many members of the target group (i.e. prosumers) as possible. Via email, over 25 organizations and individuals active in fields related to RE projects or energy cooperatives in the Netherlands were approached to diffuse the link among their members. Furthermore, the link was posted in Dutch social media communities for prosumers with a total of over 4,000 members. Next to that, the link was shared multiple times on the social media platform Twitter by several account holders working with prosumers and energy cooperatives. The exact amount of people receiving the link is difficult to determine due to the nature of the fielding process. A filtering question was used at the beginning of the survey to exclude non-prosumers, thereby ensuring that the individuals filling the survey belonged to the target group. Finally, the representativeness of the sample is limited, as solely prosumers who have access to the internet and the previously mentioned communication channels received the survey. The fielding took place over seven weeks from the 14th of April 2020 until the 1st of June 2020.

To ensure that only participants who gave valid responses were included in the results analysis, speeders who took less than 50% of the median time to finalize the survey were removed. The proposed minimum sample size for this DCE is 56[4]. Out of 90 completed surveys, four participants were classified as speeders and twelve respondents were excluded because they indicated that they did not have a renewable DER. Therefore, 74 respondents were included in the final sample for the analysis. In summary, the sample can be described as being dominated by middle-aged males with a high level of education living in smaller sized households within medium-sized cities or in rural areas, who have a relatively high household net income compared to the Dutch average of €31,000. Details can be found in Table 3. Of the surveyed prosumers, almost half (45%) were a member of an energy cooperative or community. The socio-economic characteristics of the sample are similar to those of other surveys with energy cooperative members (Bauwens, 2019) and prosumers (Hahnel et al., 2020).

Table 3. Socio-demographic characteristics of the final sample (N=74).

| Demographic | | *%* |
|---|---|---|
| Age | younger than 25 | 0 |
| | 25-35 | 4 |
| | 36-45 | 16 |
| | 46-55 | 28 |
| | 56-65 | 38 |
| | older than 65 | 14 |
| Gender | Female | 12 |
| | Male | 85 |
| | Third gender/Non-binary | 1 |
| | Prefer not to say | 1 |
| Education | No formal education | 0 |
| | High school diploma | 3 |
| | MBO (vocational) | 14 |
| | HBO (applied sciences) | 28 |
| | Bachelor's degree | 8 |
| | Master's degree | 36 |
| | PhD or higher | 9 |
| | Prefer not to say | 1 |
| Household location | Large city | 22 |
| | Medium-sized city | 35 |
| | Small town | 12 |
| | Rural community | 31 |
| Household size | 1 | 5 |

---

[4] According to the proposed rule of thumb to calculate the minimum required sample size for a DCE using the equation $N \geq 500 \cdot c \cdot t \cdot a$ (Orme, 2019). Where c equals the largest number of levels for any of the attributes, t the number of choice tasks and a is the number of alternatives. The number 500 represents the least amount of times each attribute level must be represented in the DCE to achieve sufficient stability in estimates. In this study c equals 3, t is 9 (without holdout tasks), and a is 3.

|  | 2 | 45 |
|---|---|---|
|  | 3 | 19 |
|  | 4 | 23 |
|  | 5 or more | 8 |
| Household net annual income | under €20,000 | 5 |
|  | €20,000 - €39,999 | 15 |
|  | €40,000 - €59,999 | 35 |
|  | €60,000 - €79,999 | 20 |
|  | over 80,000€ | 24 |

3.5. Data analysis

The DCE was analyzed using the Hierarchical Bayes (HB) estimation method, which is the most commonly used method of analysis for data from conjoint studies (Rossi and Allenby, 2003). The HB estimation model can be used to calculate average part-worth utilities from individual part-worth utilities of each respondent. Part-worth utilities measure the contribution of attribute levels to the overall utility, i.e. the influence of a change of the respective variable on the prosumers likelihood to participate in a specific type of electricity trading (Lüthi and Prässler, 2011).

The Lighthouse Studio software used for the calculations assumes the differences in predicted and actual choices are distributed normally and independently of one another (Sawtooth Software, 2020). Then, several thousand iterations are performed to adjust each respondent's utilities to reflect the optimal mix of individual choices and sample averages (Howell, 2009). Because part-worth utilities are interval data which were randomly scaled to an additive constant within each attribute, it is not possible to compare utility values between different attributes. Therefore, part-worth utilities are zero-centered within each attribute and the sum of the average differences between best and worst levels across all attributes is equal to the number of attributes times a hundred (Orme, 2010). Thereby, differences between attribute levels can be compared. Finally, the importance scores of each attribute can be calculated by taking the range of the attributes' utility values, i.e. the highest and the lowest part-worth utility of each attribute (Lüthi and Prässler, 2011). Here, a bigger range means that the attribute is deemed to have higher importance.

3.6. Data validity

The HB model computation was set to 100,000 iterations (of which the first 50,000 were discarded) before convergence was assumed. The values are based on a logit model that calculates the probabilities of respondents choosing tasks. The Percent Certainty[5] indicates that the log likelihood is on 72.3% between chance value and the perfect fit value (Orme, 2009). The Fit Statistic Root Likelihood[6] indicates that the model predicts the outcome of a choice task correctly 73.8% of the time (Orme, 2009).

The three included holdout tasks were analyzed as an indication for the validity of the results. The HIT rate was calculated, which is a measure to assess how well the modelled individual utilities predict actual responses. The HIT rate was identified to be at 70.27%, which indicates that the model predicts respondents' choices well (Orme, 2009). The first and second holdout task were identical to test the test-retest reliability. 81.1% of respondents selected both identical choice tasks that appeared in the third and the twelfth position of the DCE. This serves as indication that most respondents paid good attention when they chose alternatives, and therefore, for the validity and usefulness of the results. The third holdout task was used to measure convergent validity. A majority, 58.1% of the respondents, chose the alternative with environmental benefits, which is in line with findings from the utility analysis (see 4.3.).

4. Results and Discussion

4.1. Prosumers' characteristics and attitudes

Regarding the reasons why respondents installed a renewable energy system (Figure 2), the vast majority (over 85%) stated they wanted to tackle the climate change problem by being part of the clean energy transition. 75% of the respondents wanted to reduce their energy costs. The reasons of having more control over energy production and use, receiving a subsidy, and having an interest in the technology were each chosen by about a third of the participants. When asked about the reasons for choosing their current energy provider (Figure 3), the majority (72%) indicated they chose them because the supplier provided green energy, while only 20% made the decision based on the company offering the lowest costs. The most prominent reasons to become an energy cooperative member (Figure 4) was to tackle the climate change problem (82%) and to decentralize the energy production in the Netherlands (65%). This is in line with studies which find that cooperative membership is primarily driven by environmental motivations rather than financial return (Bauwens, 2017). About half of the cooperative members wanted to create a sense of local community (50%) and to improve revenues for the community (41%).

---

[5] The Percent Certainty indicates where the analysis results lie within the range of complete chance and the perfect solution. A value of 0 would mean that the model fits the data only at chance level, while a value of 1 means that the data is a perfect fit of the model.

[6] The Fit Statistic Root Likelihood is the geometric root of likelihoods across all respondent tasks. It can be interpreted as follows: Respondents have 3 alternatives to choose from. At a random guess each alternatives probability of being picked is 1/3 (33%). A value of 1 indicates perfect fit of the model.

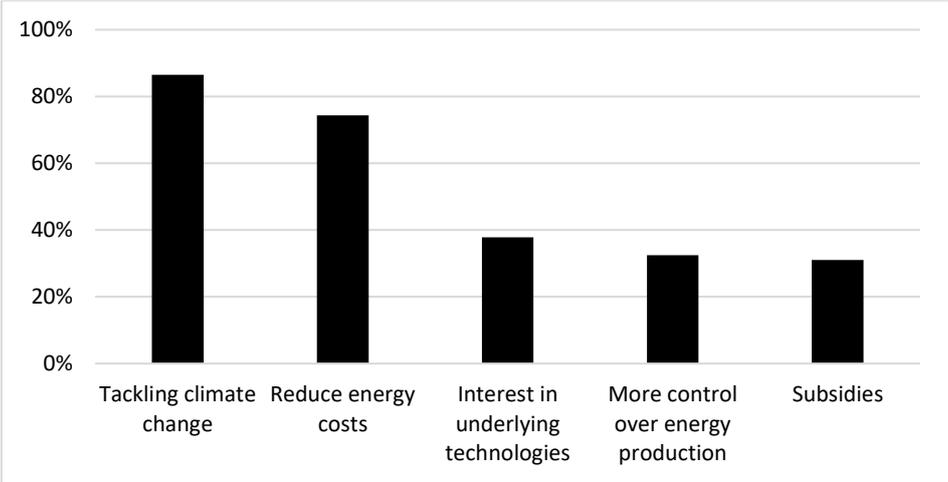

Figure 2. Reasons for installing a renewable energy system (N=74).

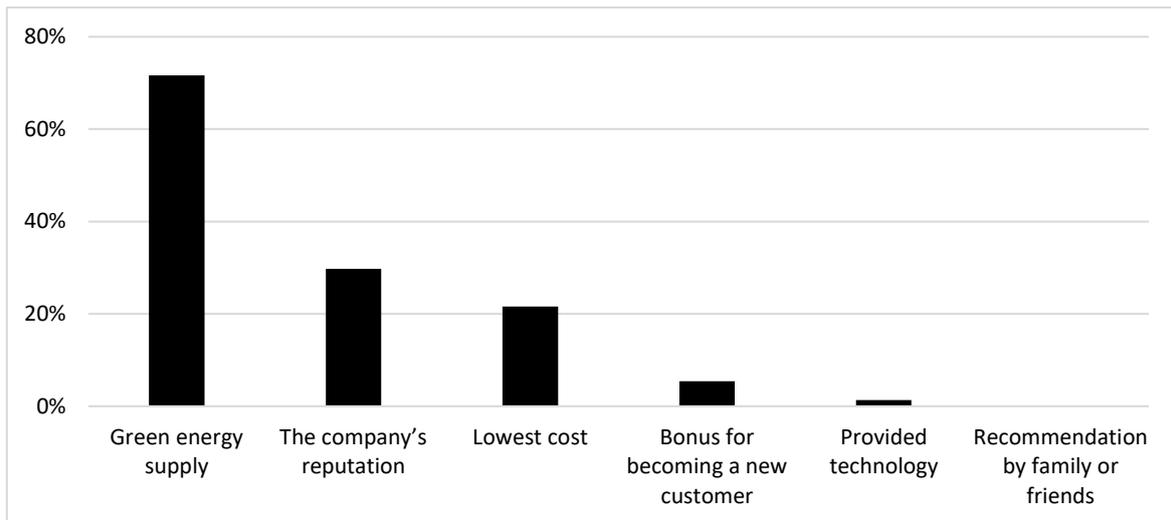

Figure 3. Reasons for choosing the current energy provider (N=74).

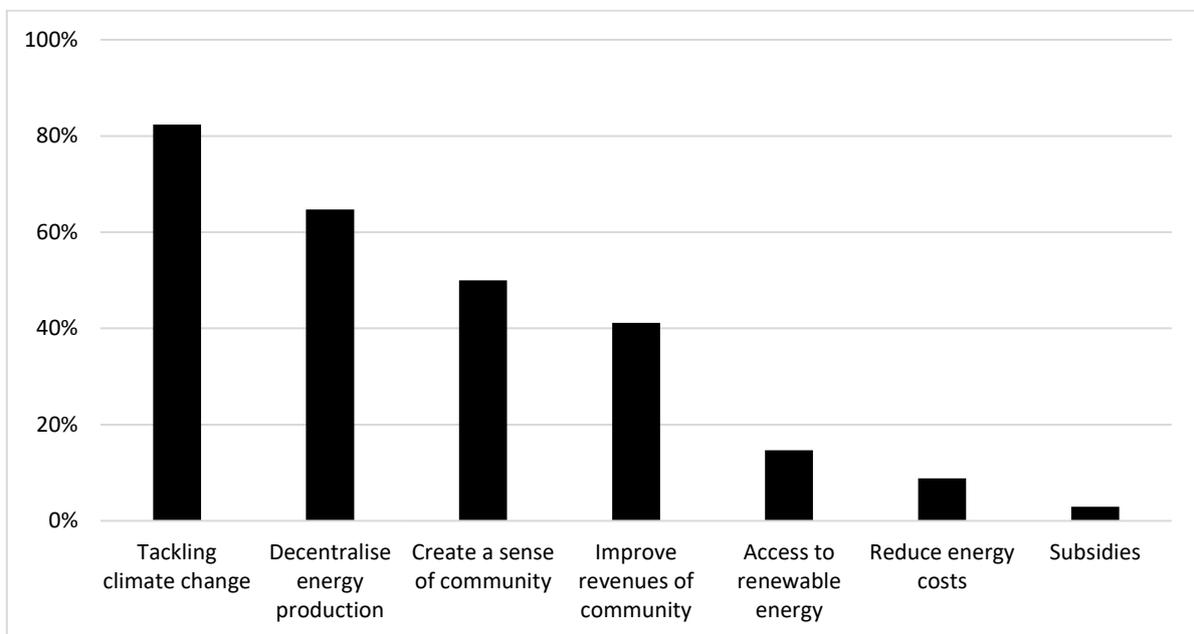

Figure 4. Reasons for becoming a member of an energy cooperative or community (N=34).

Figure 5 presents the results of the surveyed prosumers' attitudes towards the environment, their local community and technologies. The two statements measuring the construct *environmental attitude* received a high combined mean value of 4.14, indicating that most respondents have a high environmental concern. This is in line with other studies investigating prosumers' environmental attitudes (Hahnel et al., 2020) and environmental concerns being a main driver for installing a DER. It is also consistent with the previous finding that 82% of the participants became prosumers to tackle the climate change problem (see Figure 2). The two statements measuring the construct *community attitude* were assessed with lower agreement values. Their combined mean value is 3.23, while this value is somewhat higher among energy cooperative members (3.44). The last two statements which measured the construct *technological attitude* were both rated with higher agreement. The samples combined mean value for the two statements is 4.19, and thereby higher than in studies which did not specifically target prosumers (e.g. Hackbarth and Löbbe, 2020; Mengelkamp et al., 2019). The sample's high interest in technologies reflects the notion that individuals are less likely to adopt DER installations when they lack technological interest (Palm, 2018). The finding that prosumers show a higher interest in technologies can therefore influence their participation in electricity trading, as also suggested in previous studies (Mengelkamp et al., 2018; Reuter and Loock, 2017; Wilkinson et al., 2020).

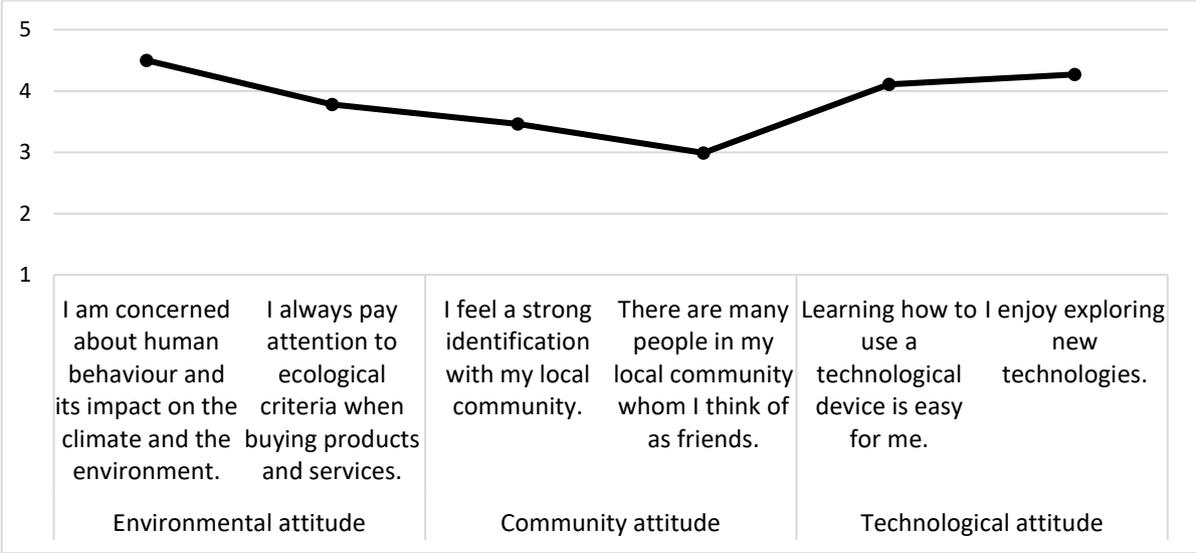

Figure 5. Results of environmental, community and technological attitude statement questions.

These attitudes were also compared between cooperative members and non-members. Significance was tested with a two-tailed t-test. Regarding the attitudes towards the environment, cooperative members had on average a higher approval of the two respective statements (M = 4.38) than non-members (M = 3.94), t(73) = 2, p = .009. This, again, shows that cooperative members are strongly driven by environmental motives. Cooperative members' attitudes towards their community (M =

3.44) were also significantly higher than those of non-members (M = 3.04), t(73) = 2, p = .026. This underlines that cooperative members feel a stronger sense of belonging in their local community than individual prosumers. As both these aspects were identified as drivers for participation in P2P trading (Mengelkamp et al., 2018), this suggests that members of energy cooperatives are especially likely to participate in P2P electricity trading.

4.2. Willingness to exchange for free or indirect financial returns

Regarding the willingness to exchange surplus electricity without a return or for an indirect financial return (Figure 6), our findings show that 60% of respondents would be willing to give surplus electricity for free and 76% would be willing to do so for an indirect financial return. In addition, 77% of the participants who would give away surplus for free and 70% of those who would do so for an indirect monetary return, would give it to a household that cannot afford electricity. These findings support the idea that prosumers see electricity as social capital they give to consumers in need (Jogunola et al., 2017) and supports the previous finding that participants of P2P electricity platforms show a high interest in social equity (Wilkinson et al., 2020).

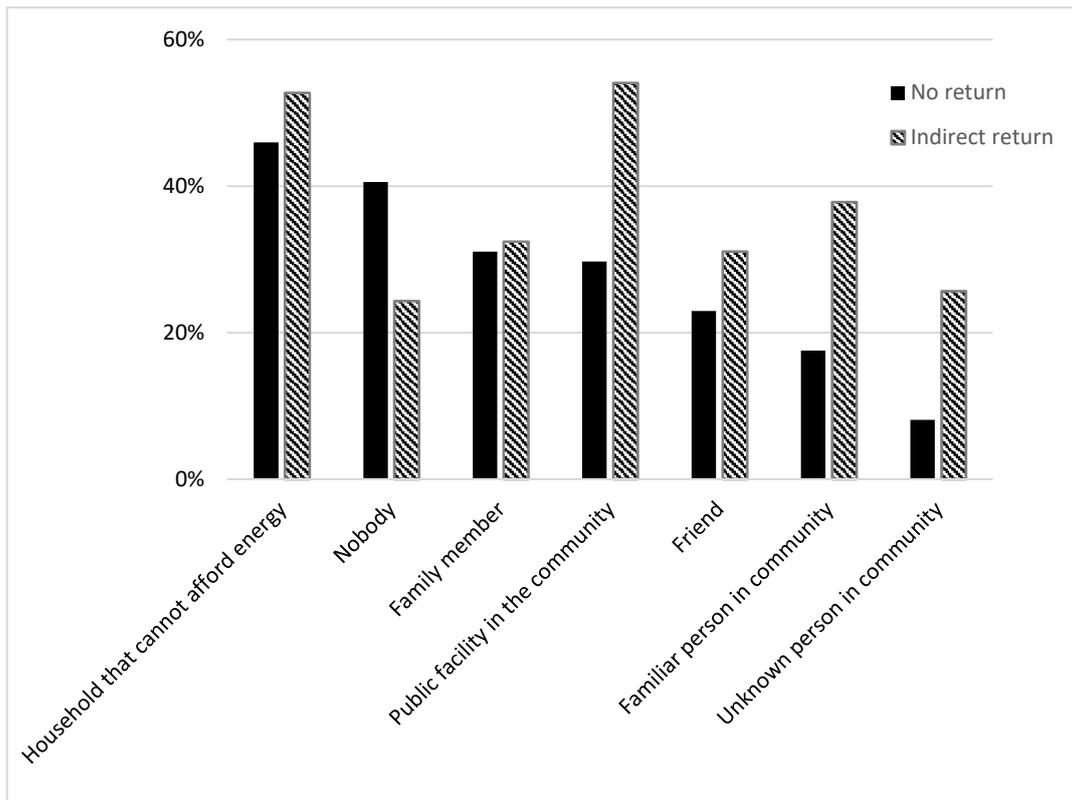

Figure 6. Results for the willingness to give surplus electricity for none or an indirect return (N=74). Note: Since participants were given the option to select multiple responses, percentages may add up to over 100%.

Moreover, of the participants who would give away surplus for free, 52% would give it to a family member, 50% to a public facility (e.g. schools) in the community, 39% to a friend, 30% to someone in their community they know and 14% to someone they do not know. Of the participants who would give away surplus for an indirect monetary return, 72% would give it to a public facility, 50% to someone in their community they know, 43% to a family member, 41% to a friend and 34% to someone they do not know. These findings indicate that non-monetary compensation is more likely to be accepted with family members, friends, and familiar community members. This is consistent with Singh et al.'s (2018) study in the context of off-grid energy communities in rural India, which shows that the willingness to accept returns different from in-cash payments increases with the strength of the social relationship between energy providers and consumers. Our finding supports this result in the context of a high-income country in the global north. It also underscores the importance of community norms in facilitating P2P electricity trading.

In addition, when comparing prosumers participating in the study who were energy cooperative members to those who were not, our findings show that cooperative members are more willing to exchange surplus electricity for free (68% vs. 53%) and for indirect financial returns (85% vs. 68%) than non-members. This is consistent with the cooperative ideal, according to which energy cooperatives strive to democratize energy production and consumption in a more equal way, promoting solidarity among members (e.g. Vansintjan, 2015). This result also illustrates the potential key role that energy cooperatives may play in advancing a socially just energy transition.

4.3. Prosumer preferences for electricity trading

Using the Hierarchical Bayes estimation method, the part-worth utilities with standard deviations of the attribute levels were calculated (Table 4)[7]. Within the attribute *$CO_2$ emissions,* the level "low" has the highest utility and is therefore preferred, and the level "high" has the lowest utility. For the attribute *selling price*, the highest price of 20 €ct/kWh has the highest utility, and the lowest price of 10 €ct/kWh has the lowest utility. Having the choice of the *social connection with an electricity trading partner*, on average respondents preferred someone they know closely (e.g. a friend or a relative), which was indicated by the level "close" having a higher utility as compared to the two other levels of the attribute. When it comes to the attribute *additional effort (time)*, no additional amount of time spent per month was preferred, while an additional 4 h/month had the lowest utility score. The *improved efficiency* level of 30% had a superior utility score, compared to the ones of the levels 15% and 0%. Finally, the attribute *self-sufficiency* had the highest utility value for the level "high" and the level "low" the lowest utility.

---

[7] The part-worth utilities were zero-centred, i.e. normalised, in order to facilitate their interpretation. The higher the utility score of an attribute level is, the more attractive it has been on average for respondents in the study. A negative utility score does not imply that the attribute level is unattractive, but solely that it is less attractive relative to other levels. An average utility score is built from the individual utility scores of each respondent and for each attribute level. The average utility scores cannot be compared across attributes but only across levels within the same attribute.

Table 4. Average utility scores and standard deviations.

| Attribute | Attribute levels | Average Utilities | Standard deviation |
|---|---|---|---|
| $CO_2$ emissions | Low | 79.9 | 83.39 |
|  | Medium | 32.14 | 19.28 |
|  | High | -112.04 | 80.89 |
| Selling price | 10 €ct/kWh | -25.12 | 37.8 |
|  | 15 €ct/kWh | -2.26 | 17.36 |
|  | 20 €ct/kWh | 27.38 | 41.27 |
| Social connection | None | -20.44 | 19.08 |
|  | Direct | 1.74 | 19.44 |
|  | Close | 18.7 | 19.84 |
| Additional effort (time) | 0 h/month | 45.41 | 49.39 |
|  | 2 h/month | 9.54 | 25.48 |
|  | 4 h/month | -54.95 | 50.57 |
| Improved efficiency | 0% | -29.37 | 24.42 |
|  | 15% | 6.98 | 24.83 |
|  | 30% | 22.39 | 22.87 |
| Self-sufficiency | Low | -17.85 | 34.08 |
|  | Medium | -7.52 | 14.44 |
|  | High | 25.38 | 29.21 |

The HB model also estimates the attributes importance scores, which indicate the importance of an attribute relative to the other attributes. Figure 7 gives a graphic representation. With 37.74%, the attribute *$CO_2$ emissions* had by far the highest impact on respondents' decision-making process. This is followed by the attribute *additional effort (time)* with 19.60%. After this the relative importance of the remaining four attributes *selling price, improved efficiency, self-sufficiency* and *social connection* is close together and appears in this order while ranging from 12.58% to 8.53%.

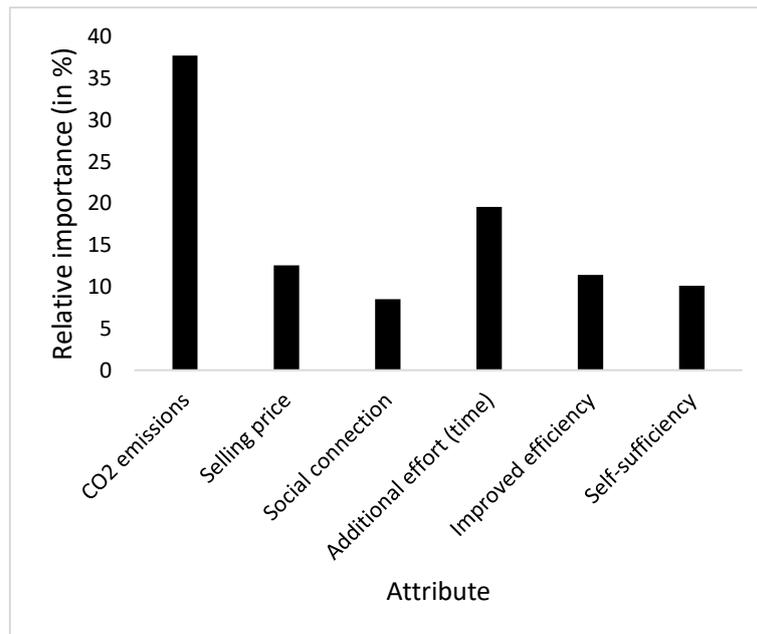

Figure 7. Average Importance scores

These results show that the average respondent's choices are mostly influenced by the attribute $CO_2$ emissions, while the other five attributes influence the decisions to a lesser extent. This implies that respondents would make trade-offs, for example accept a lower selling price, to ensure that their electricity use is associated with low $CO_2$ emissions. This is in line with previous findings that showed that a positive attitude towards the environment is the main predictor for the willingness to participate in P2P electricity trading (Hackbarth and Löbbe, 2020; Mengelkamp et al., 2018). This result also reflects the more nuanced approach of targeting prosumers instead of general electricity consumers, for whom other studies identified economic attributes to be most influential (Hahnel et al., 2020; Mengelkamp et al., 2019). In the context of this survey, the dominance of the environmental attribute may be explained by the sample consisting of RE prosumers, a large share of whom are also members of an energy cooperative. They are thus particularly strongly committed to the development of sustainable energy. This can also be related to the finding that a large majority of respondents installed a renewable DER and joined an energy cooperative primarily to tackle the climate change problem by being part of the clean energy transition, which further confirms that they are particularly strongly environmentally concerned.

The second priority of respondents when choosing an electricity trading concept was the amount of additional effort (in terms of time spent monthly) that would be potentially required for managing the trading processes. Prosumers were more reluctant to choose trading scenarios that included more additional effort. This was particularly the case for respondents which were identified to have lower interest in new technologies and stated they had difficulties when using them. This finding supports similar tendencies which were found in other studies where an increasing need for coordination was identified as a potential risk for the adoption of P2P electricity trading (Reuter and Loock, 2017) and where systems with easy implementation and comfortable service bundles were preferred (Hackbarth and Löbbe, 2020). This implies that a well-managed and highly automated trading platform is vital for the success of P2P trading.

The price for which surplus electricity can be sold, on the other hand, played a significantly lower role in the decision-making of respondents compared to the two previously mentioned attributes. Still, when choosing a trading scenario, prosumers preferred higher over lower selling prices, which is in line with findings from other studies (Hahnel et al., 2020; Mengelkamp et al., 2019). This also corresponds to a majority of respondents having stated that they installed the renewable DER in order to reduce their energy costs. While the economic benefits of P2P trading should be emphasized, they may not play the most significant role in prosumers decision to participate in P2P trading.

The next finding concerns the extent to which respondents found it important that their choice of electricity trading influences the overall efficiency of electricity supply. While preferring systems with the highest possible efficiency improvement, respondents' choices were not heavily influenced by this attribute. The role of efficiency improvements for individuals' decisions has not been investigated before in the context of P2P electricity trading. Yet, it is one of the benefits over large-scale centralized electricity grids (Jogunola et al., 2017). Therefore, the communication of this aspect may be beneficial to achieve increasing participation, although it was not seen as one of the most important aspects. Especially so, because the improved efficiency has positive impacts on both the $CO_2$ emissions and the selling price by utilizing more electricity from the prosumers' renewable DER.

Although trading scenarios with high levels of self-sufficiency were preferred by respondents, their decisions were not highly influenced by the degree of self-sufficiency presented. This corresponds with the additional finding that only a minority of respondents installed their renewable DER to have more control over their own energy production and use. In addition, this matches statements given in the interviews with prosumers that preceded the survey, in which many interviewees mentioned that they expect to remain reliant on large centralized grids and thereby big energy providers. Still, some interviewees expressed that self-sufficiency is an aspect that they deem important in the light of insecurities of future electricity supply related to the phase-out of electricity from fossil sources in the Netherlands. This dichotomous finding both supports and contradicts literature that ascertained that increased autarky, which is related to the concept of self-sufficiency, negatively influences individuals' willingness to participate in P2P electricity trading (Ecker et al., 2018). Nevertheless, due to the variability of energy supply from DERs, trading ones' surplus electricity with peers can be a means to achieve a certain level of self-sufficiency from larger energy companies through local electricity exchanges, without compromising the self-sufficiency of prosumers since only surplus electricity is traded. A further distinction should be made between self-sufficiency on the individual level and on the community level. While P2P trading increases the latter, prosumers become more dependent on their community in the process. The preferences of prosumers regarding these different forms of self-sufficiency should be addressed in future research.

Even though self-sufficiency does not seem to be as important to respondents as other aspects of electricity trading, its impact should be investigated again when prosumers are confronted with potential future insecurities of electricity supply and remuneration schemes. Additionally, future research may investigate explicitly how prosumers' desire for control and their trust in existing energy infrastructure, influences their preference for electricity trading schemes. These two aspects were indirectly addressed within the attribute *self-sufficiency*, which was explained as entailing the level of autonomy from energy providers and the level of data privacy in the experiment (see Appendix A, Table A.2). To further differentiate between their individual influence on prosumers' trading preferences, future research may explicitly investigate the two aspects of independence in terms of energy use and in terms of data privacy separately.

The ability to choose electricity trading partners according to one's preferences is a unique feature of P2P electricity trading. Although having the smallest impact on the choices made by the participating prosumers, a valuable finding was that respondents preferred having a trading partner they stand in a close social relationship with, i.e. a family member or friend. This contradicts findings by Reuter and Loock (2017), who found no indication for preferred relations to trading partners.

The previous analyses were repeated separately for cooperative members and non-members to identify potential differences. Significance was tested using a two-sided t-test to compare the means of utilities and importance scores. The average utility scores for the attribute levels 10 €ct/kWh and 20 €ct/kWh within the attribute *selling price* differed significantly between members and non-members. Energy cooperative members' average utility was significantly higher (M = -12.76 vs. M = -35.62) for the lower price and significantly lower for the higher price (M = 13.01 vs. M = 39.59), t(73) = 2, p = .007 and p = .003. This result indicates that members, on average, have less extreme preferences for the amount of money they can sell their surplus electricity for. This is also indicated by a significantly lower importance score (M = 8.84% vs. M = 15.77%) for the attribute *selling price*, t(73) = 2 , p = .001, meaning that the attribute had a lower impact on energy cooperative members' decision-making process, compared to non-members. This result suggests that energy cooperative members, on average, are less motivated by economic factors than non-members, in accordance with previous studies which show that cooperative members are primarily motivated by environmental or other non-financial aspects (Bauwens, 2017; Radtke, 2014).

## 5. Conclusion and policy implications

This research was based on a survey which included a DCE conducted with prosumers in the Netherlands to elicit their preferences regarding P2P electricity trading. The findings show that the environmental attribute ($CO_2$ emissions) was by far the most important factor influencing respondents' decisions about electricity trading concepts. The second most influential attribute was additional effort (time), while the economic attribute (selling price), which was expected to play a major role in the decision making, only came third and almost on the same level as the remaining three attributes: improved efficiency, self-sufficiency, and social connection. This shows that the decision about partaking in P2P electricity trading among the participating prosumers is not solely financially driven, but highly influenced by the environmental impact of the consumed electricity and the extra efforts required.

Furthermore, the study extends the current state of research on P2P electricity trading by investigating prosumers' willingness to provide surplus electricity for free or indirect financial returns. More than half of the respondents indicated that they would provide free electricity to certain groups, and three quarters would accept an indirect return. The actors they would most likely give electricity to under these conditions are households that cannot afford energy or public facilities in their community. This result is especially important for advancing a socially fair energy transition, as it illustrates how P2P trading could help low-income households gain access to renewable electricity through their local community.

In addition, the results demonstrate that energy cooperative members attach less importance to the attribute selling price and are more likely to choose trading scenarios with lower selling prices, suggesting that their participation in P2P electricity trading is more strongly driven by non-economic factors. Cooperative members also show a higher concern for their community and feel more strongly connected to it. Finally, they were more willing to participate in electricity exchanges which entailed no return or an indirect financial compensation. This indicates that next to the individual economic benefits, prosumers are also interested in collectivizing economic benefits within their community.

Admittedly, this study also has limitations. First, it must be noted that the survey had a rather small sample size of 74 respondents, whereas 68% of comparable studies using a DCE have sample sizes of over 100 (Bekker-Grob et al., 2015). Reasons for the small number of participants are the limited resources available for the study, the decision to only include prosumers and the used distribution channels. Nevertheless, significance tests showed that enough datasets were available to receive reliable DCE results, which comes from the fact that each participant answered nine choice tasks, resulting in a total of 666 datasets. Still, a potential consequence of the small sample size is a reduced generalizability of the study from the rather homogenous socio-economic characteristics of the respondents. Hence, it cannot be ruled out that the study's findings are not applicable to the general population of Dutch prosumers. To be able to derive geographically sensitive policies, we recommend repeating adapted versions of the study on regional levels, e.g. on municipality level, with a sample that is representative of the specific region.

Second, one can expect that a participant's intention alone does not necessarily lead to the corresponding behavior, i.e. in this case the participation in a P2P electricity market. This so-called intention–behavior gap describes the discrepancies between an individual's intentions regarding their actual behaviors, which can be explained with too optimistic goals and missing abilities and resources to achieve these goals (Sheeran and Webb, 2016). The proposed survey method, i.e. the DCE, has potential to work around this bias, as it does not directly ask for participants' intentions, but implicitly investigates their preferences. To investigate the impact of intention-behavior gaps, further research could conduct longitudinal studies on this matter, where participants' indicated preferences are checked against actual behavior.

Third, the criterion for survey participants to be included in the final sample was to be a prosumer. This criterion was fulfilled when a participant indicated that they own a DER. Since there was not additional measure to verify the criterion, the results are reliant on the participant truthfully filling in the questionnaire. Next to prosumers, consumers may also participate in P2P electricity trading. Therefore, we recommend repeating the study with a larger sample that also includes consumers in order to elicit their trading preferences and investigate potential distinctions to prosumers.

Following the findings of this study, several implications for the marketing of P2P electricity trading, the platform design and required policies can be given. When marketing a P2P electricity trading platform to existing prosumers, operators should emphasize the environmental benefits of P2P electricity trading to attract the large group of environmentally driven prosumers. Economically driven prosumer can be attracted by communicating that P2P trading can result in higher selling prices than conventionally feeding electricity into a central grid. To attract niche market customers that attach high value to community aspects and social exchange, the unique characteristics of P2P trading (e.g. the possibility to exchange electricity with selected peers) should be communicated. In addition, peer effects within local communities (e.g. word of mouth) can be utilized to increase the awareness for P2P trading. While the design of a P2P trading platform should incorporate options to set transactions according to individual preferences, the platform should, at the same time, enable automatic transactions which require minimal effort (e.g. by incorporating an intelligent agent system that enables automatic electricity transaction), thereby addressing the concerns of participants who highly value electricity trading with little additional effort.

Furthermore, policy makers can take into consideration that most prosumers were willing to provide generated surplus electricity for a non-monetary return or even for no return at all to drive the renewable energy transition in a fairer and just way. Within P2P electricity trading, the exchange of more affordable renewable energy with low-income households can be promoted as a way to spread the use of low-carbon energy among socio-demographic groups who are otherwise excluded from it. Because energy cooperative members in particular showed a higher willingness to participate in these exchanges, energy cooperatives can adopt P2P electricity trading concepts to emphasize their role as key actors in a fair energy transition.

More generally, energy cooperatives, as they are strongly rooted in local communities and promote the values of fairness and solidarity among their members and with the rest of society, can act as forerunners and facilitators for a socially just development of P2P electricity trading. Hence, they should be involved in the process of setting-up P2P trading as initiators and governing institution of the P2P electricity market (Reuter and Loock, 2017). This is especially important in the context of changing support mechanisms for renewable energy, which often force local organizations such as energy cooperatives to search for new financially sustainable business models and to diversify their activity portfolio (Bauwens, 2020; Herbes et al., 2017). In this perspective, P2P electricity trading appears to be a particularly attractive model for them. However, next to the technical requirements, the incorporation of P2P electricity trading schemes into cooperatives' business models also requires a certain open-mindedness of cooperative members. This may be facilitated through an open exchange between cooperative members and the P2P electricity network organizers.

Finally, a set of clear policies and guidelines for P2P electricity trading on national levels are required to reduce insecurities regarding the regulatory framework for P2P electricity trading (e.g. regarding the taxation of any potential profit made from selling surplus electricity). Currently, many EU countries allow self-trading of electricity on a very limited scale or solely indirectly through existing electricity market actors and their network (van Soest, 2018). Yet, the complementary feature of P2P trading to empower prosumers and consumers by becoming independent from actors with large market power can only be realized when prosumer self-trading is acknowledged in regulatory policy.

Overall, this study demonstrates that P2P electricity trading schemes incorporate many elements that are considered important and are valued by prosumers in the Netherlands. This finding, together with the future policy changes in the Dutch net-metering system and an increasing adoption of renewable DER and digitalization by households, supports the concept of P2P electricity trading as a new way of empowering prosumers and driving the renewable energy transition in the Netherlands. It is hoped that this study will encourage future research bringing additional elements of answers to these fascinating questions and illuminating further the roles of P2P electricity trading in transitioning towards just and sustainable energy systems.

## Declaration of competing interests

The authors declare that they have no known competing financial interests or personal relationships that could have appeared to influence the work reported in this paper.

## Acknowledgements

This work is part of the research project: "A Blockchain-based platform for peer-to-peer energy transactions between Distributed Energy Resources (B-DER)", which received funding from the Netherlands Enterprise Agency (RVO) within the Topsector Energy framework, project number: 1621404. The authors want to thank Dr. Esther Mengelkamp for sharing her expertise on local electricity markets and conducting choice experiments in the early stages of this research. We also want to thank Dr. Hendrik Kondziella of Leipzig University for providing feedback on the first version of this article. Finally, we thank Sawtooth Software Inc. for sponsoring this research with an academic grant to use the software.

# References

Akorede, M.F., Hizam, H., Pouresmaeil, E., 2010. Distributed energy resources and benefits to the environment. Renewable Sustainable Energy Rev. 14, 724–734. 10.1016/j.rser.2009.10.025.

AlSkaif, T., Crespo-Vazquez, J.L., Sekuloski, M., van Leeuwen, G., Catalao, J.P.S., 2021. Blockchain-based Fully Peer-to-Peer Energy Trading Strategies for Residential Energy Systems. IEEE Trans. Ind. Inf., 1. 10.1109/TII.2021.3077008.

AlSkaif, T., Zapata, M.G., Bellalta, B., Nilsson, A., 2017. A distributed power sharing framework among households in microgrids: A repeated game approach. Computing 99, 23–37. 10.1007/s00607-016-0504-y.

Axsen, J., Orlebar, C., Skippon, S., 2013. Social influence and consumer preference formation for pro-environmental technology: The case of a U.K. workplace electric-vehicle study. Ecol. Econ. 95, 96–107. 10.1016/j.ecolecon.2013.08.009.

Balcombe, P., Rigby, D., Azapagic, A., 2013. Motivations and barriers associated with adopting microgeneration energy technologies in the UK. Renewable Sustainable Energy Rev. 22, 655–666. 10.1016/j.rser.2013.02.012.

Balcombe, P., Rigby, D., Azapagic, A., 2014. Investigating the importance of motivations and barriers related to microgeneration uptake in the UK. Appl. Energy 130, 403–418. 10.1016/j.apenergy.2014.05.047.

Bauwens, T., 2016. Explaining the diversity of motivations behind community renewable energy. Energy Policy 93, 278–290. 10.1016/j.enpol.2016.03.017.

Bauwens, T., 2017. Designing Institutions for Collective Energy Action: The Roles of Renewable Energy Cooperatives in a Polycentric Low-Carbon Transition. Unpublished doctoral dissertation., Liège.

Bauwens, T., 2019. Analyzing the determinants of the size of investments by community renewable energy members: Findings and policy implications from Flanders. Energy Policy 129, 841–852. 10.1016/j.enpol.2019.02.067.

Bauwens, T., 2020. When community meets finance. Nat. Energy 5, 119–120. 10.1038/s41560-019-0547-3.

Bauwens, T., Gotchev, B., Holstenkamp, L., 2016. What drives the development of community energy in Europe?: The case of wind power cooperatives. Energy Res. Soc. Sci. 13, 136–147. 10.1016/j.erss.2015.12.016.

Bekker-Grob, E.W. de, Donkers, B., Jonker, M.F., Stolk, E.A., 2015. Sample Size Requirements for Discrete-Choice Experiments in Healthcare: A Practical Guide. Patient 8, 373–384. 10.1007/s40271-015-0118-z.

Bollinger, B., Gillingham, K., 2012. Peer Effects in the Diffusion of Solar Photovoltaic Panels. Market. Sci. 31, 900–912. 10.1287/mksc.1120.0727.

Brinkel, N.B.G., Gerritsma, M.K., AlSkaif, T.A., Lampropoulos, I., van Voorden, A.M., Fidder, H.A., van Sark, W.G.J.H.M., 2020. Impact of rapid PV fluctuations on power quality in the low-voltage grid and mitigation strategies using electric vehicles. Int. J. Electr. Power Energy Syst. 118, 105741. 10.1016/j.ijepes.2019.105741.

Buth, M.C., Wieczorek, A.J., Verbong, G.P.J., 2019. The promise of peer-to-peer trading?: The potential impact of blockchain on the actor configuration in the Dutch electricity system. Energy Res. Soc. Sci. 53, 194–205. 10.1016/j.erss.2019.02.021.

Campos, I., Pontes Luz, G., Marín-González, E., Gährs, S., Hall, S., Holstenkamp, L., 2020. Regulatory challenges and opportunities for collective renewable energy prosumers in the EU. Energy Policy 138, 111212. 10.1016/j.enpol.2019.111212.

Claudy, M.C., Peterson, M., O'Driscoll, A., 2014. Understanding the Attitude-Behavior Gap for Renewable Energy Systems Using Behavioral Reasoning Theory. J. Macromarketing 33, 273–287. 10.1177/0276146713481605.

Crespo-Vazquez, J.L., AlSkaif, T., Gonzalez-Rueda, A.M., Gibescu, M., 2020. A Community-Based Energy Market Design Using Decentralized Decision-Making under Uncertainty. IEEE Trans. Smart Grid, 1. 10.1109/TSG.2020.3036915.

Ecker, F., Spada, H., Hahnel, U.J.J., 2018. Independence without control: Autarky outperforms autonomy benefits in the adoption of private energy storage systems. Energy Policy 122, 214–228. 10.1016/j.enpol.2018.07.028.

ECoop, 2020. Postcoderoosregeling - Postcoderoosregeling / Regeling verlaagd tarief. https://www.postcoderoosregeling.nl/. Accessed 12 May 2020.

Enlund, T., Eriksson, E., 2016. Förnybar energi för alla. Slutrapport Green Leap, KTH. KTH, Green Leap, Stockholm.

Giotitsas, C., Pazaitis, A., Kostakis, V., 2015. A peer-to-peer approach to energy production. Technol. Soc. 42, 28–38. 10.1016/j.techsoc.2015.02.002.

Hackbarth, A., Löbbe, S., 2020. Attitudes, preferences, and intentions of German households concerning participation in peer-to-peer electricity trading. Energy Policy 138, 111238. 10.1016/j.enpol.2020.111238.

Hahnel, U.J.J., Herberz, M., Pena-Bello, A., Parra, D., Brosch, T., 2020. Becoming prosumer: Revealing trading preferences and decision-making strategies in peer-to-peer energy communities. Energy Policy 137, 111098. 10.1016/j.enpol.2019.111098.

Hanson, K., McPake, B., Nakamba, P., Archard, L., 2005. Preferences for hospital quality in Zambia: Results from a discrete choice experiment. Health Econ. 14, 687–701. 10.1002/hec.959.

Heckman, J.J., 2015. Introduction to a Theory of the Allocation of Time by Gary Becker. Econ. J. 125, 403–409. 10.1111/ecoj.12228.

Herbes, C., Brummer, V., Rognli, J., Blazejewski, S., Gericke, N., 2017. Responding to policy change: New business models for renewable energy cooperatives – Barriers perceived by cooperatives' members. Energy Policy 109, 82–95. 10.1016/j.enpol.2017.06.051.

HIER opgewekt, RVO, 2019. Lokale Energie Monitor 2019. https://www.hieropgewekt.nl/lokale-energie-monitor. Accessed 12 May 2020.

Howell, J., 2009. CBC/HB For Beginners. Sawtooth Software Technical Paper Series.

Jogunola, O., Ikpehai, A., Anoh, K., Adebisi, B., Hammoudeh, M., Son, S.-Y., Harris, G., 2017. State-Of-The-Art and Prospects for Peer-To-Peer Transaction-Based Energy System. Energies 10, 2106. 10.3390/en10122106.

Kalkbrenner, B.J., Roosen, J., 2016. Citizens' willingness to participate in local renewable energy projects: The role of community and trust in Germany. Energy Res. Soc. Sci. 13, 60–70. 10.1016/j.erss.2015.12.006.

Karrer, K., Glaser, C., Clemens, C., Bruder, C., 2009. Technikaffinität erfassen – der Fragebogen TA-EG, in: Lichtenstein, A., Stel, C., Clemens, C. (Ed.), Der Mensch Als Mittelpunkt Technischer Systeme. 8. Berliner Werkstatt Mensch-Maschine-Systeme. VDI Verlag, Düsseldorf, Germany, pp. 196–201.

Kooij, H.-J., Oteman, M., Veenman, S., Sperling, K., Magnusson, D., Palm, J., Hvelplund, F., 2018. Between grassroots and treetops: Community power and institutional dependence in the renewable energy sector in Denmark, Sweden and the Netherlands. Energy Res. Soc. Sci. 37, 52–64. 10.1016/j.erss.2017.09.019.

Kuckartz, U., 2000. Umweltbewußtsein in Deutschland 2000: Ergebnisse einer repräsentativen Bevölkerungsumfrage; Umweltforschungsplan des Bundesministeriums für Umwelt, Naturschutz und Reaktorsicherheit, Förderkennzeichen 299 11 132.

Lüth, A., Zepter, J.M., Crespo del Granado, P., Egging, R., 2018. Local electricity market designs for peer-to-peer trading: The role of battery flexibility. Appl. Energy 229, 1233–1243. 10.1016/J.APENERGY.2018.08.004.

Lüthi, S., Prässler, T., 2011. Analyzing policy support instruments and regulatory risk factors for wind energy deployment—A developers' perspective. Energy Policy 39, 4876–4892. 10.1016/j.enpol.2011.06.029.

Mangham, L.J., Hanson, K., McPake, B., 2009. How to do (or not to do) … Designing a discrete choice experiment for application in a low-income country. Health Policy Plan. 24, 151–158. 10.1093/heapol/czn047.

Mengelkamp, E., Schönland, T., Huber, J., Weinhardt, C., 2019. The value of local electricity - A choice experiment among German residential customers. Energy Policy 130, 294–303. 10.1016/j.enpol.2019.04.008.

Mengelkamp, E., Staudt, P., Garttner, J., Weinhardt, C., Huber, J., 2018. Quantifying Factors for Participation in Local Electricity Markets. 2018 15th International Conference on the European Energy Market, 1–5.

Ministry of Economic Affairs and Climate Policy, 2019. Integrated National Energy and Climate Plan 2021-2030: The Netherlands. https://ec.europa.eu/energy/sites/ener/files/documents/nl_final_necp_main_en.pdf. Accessed 29 October 2020.

Morstyn, T., Farrell, N., Darby, S.J., McCulloch, M.D., 2018. Using peer-to-peer energy-trading platforms to incentivize prosumers to form federated power plants. Nat. Energy 3, 94–101. 10.1038/s41560-017-0075-y.

Morstyn, T., McCulloch, M.D., 2019. Multiclass Energy Management for Peer-to-Peer Energy Trading Driven by Prosumer Preferences. IEEE Trans. Power Syst. 34, 4005–4014. 10.1109/TPWRS.2018.2834472.

Orme, B., 2009. CBC/HB v5: Software for Hierarchical Bayes Estimation for CBC Data. Sawtooth Software Technical Paper Series.

Orme, B., 2010. SSI Web v8.3: Software for Web Interviewing and Conjoint Analysis. Sawtooth Software Technical Paper Series.

Orme, B., 2019. Sample Size Issues for Conjoint Analysis. Sawtooth Software Technical Paper Series.

Palm, A., 2017. Peer effects in residential solar photovoltaics adoption—A mixed methods study of Swedish users. Energy Res. Soc. Sci. 26, 1–10. 10.1016/j.erss.2017.01.008.

Palm, J., 2018. Household installation of solar panels – Motives and barriers in a 10-year perspective. Energy Policy 113, 1–8. 10.1016/j.enpol.2017.10.047.

Palm, J., Tengvard, M., 2017. Motives for and barriers to household adoption of small-scale production of electricity: Examples from Sweden. Sustainability: Science, Practice and Policy 7, 6–15. 10.1080/15487733.2011.11908061.

Parag, Y., Sovacool, B.K., 2016. Electricity market design for the prosumer era. Nat. Energy 1, 329. 10.1038/nenergy.2016.32.

Peterson, R.A., 1994. A Meta-Analysis of Cronbach's Coefficient Alpha. J. Conusm. Res. 21, 381. 10.1086/209405.

Radtke, J., 2014. A closer look inside collaborative action: Civic engagement and participation in community energy initiatives. PPP online 8, 235–248. 10.3351/ppp.0008.0003.0008.

Reuter, E., Loock, M., 2017. Empowering Local Electricity Markets: A survey study from Switzerland, Norway, Spain and Germany. University St. Gallen, 24 pp.

Rijksoverheid, 2019. Salderingsregeling verlengd tot 2023. https://www.rijksoverheid.nl/actueel/nieuws/2019/04/26/salderingsregeling-verlengd-tot-2023. Accessed 11 February 2020.

Rommel, K., Sagebiel, J., 2017. Preferences for micro-cogeneration in Germany: Policy implications for grid expansion from a discrete choice experiment. Appl. Energy 206, 612–622. 10.1016/j.apenergy.2017.08.216.

Rossi, P.E., Allenby, G.M., 2003. Bayesian Statistics and Marketing. Market. Sci. 22, 304–328. 10.1287/mksc.22.3.304.17739.

Ruotsalainen, J., Karjalainen, J., Child, M., Heinonen, S., 2017. Culture, values, lifestyles, and power in energy futures: A critical peer-to-peer vision for renewable energy. Energy Res. Soc. Sci. 34, 231–239. 10.1016/j.erss.2017.08.001.

Sawtooth Software, 2020. Lighthouse Studio 9.8.1.

Schaffer, A.J., Brun, S., 2015. Beyond the sun—Socioeconomic drivers of the adoption of small-scale photovoltaic installations in Germany. Energy Res. Soc. Sci. 10, 220–227. 10.1016/j.erss.2015.06.010.

Seyfang, G., Park, J.J., Smith, A., 2013. A thousand flowers blooming?: An examination of community energy in the UK. Energy Policy 61, 977–989. 10.1016/j.enpol.2013.06.030.

Sheeran, P., Webb, T.L., 2016. The Intention-Behavior Gap. Soc. Personal. Psychol. Compass 10, 503–518. 10.1111/spc3.12265.

Singh, A., Strating, A.T., Romero Herrera, N.A., Mahato, D., Keyson, D.V., van Dijk, H.W., 2018. Exploring peer-to-peer returns in off-grid renewable energy systems in rural India: An anthropological perspective on local energy sharing and trading. Energy Res. Soc. Sci. 46, 194–213. 10.1016/j.erss.2018.07.021.

Tajfel, H.E., 1978. Differentiation between social groups: Studies in the social psychology of intergroup relations. Academic Press.

United Nations, 2016. Paris Agreement. http://unfccc.int/files/essential_background/convention/application/pdf/english_paris_agreement.pdf. Accessed 13 February 2020.

van Leeuwen, G., AlSkaif, T., Gibescu, M., van Sark, W., 2020. An integrated blockchain-based energy management platform with bilateral trading for microgrid communities. Appl. Energy 263, 114613. 10.1016/j.apenergy.2020.114613.

van Soest, H., 2018. Peer-to-peer electricity trading: A review of the legal context. Compet. Regul. Netw. Ind. 19, 180–199. 10.1177/1783591719834902.

Vansintjan, D., 2015. The energy transition to energy democracy. REScoop.eu, Antwerp, Belgium.

Vasseur, V., Kemp, R., 2015. The adoption of PV in the Netherlands: A statistical analysis of adoption factors. Renewable Sustainable Energy Rev. 41, 483–494. 10.1016/j.rser.2014.08.020.

Wilkinson, S., Hojckova, K., Eon, C., Morrison, G.M., Sandén, B., 2020. Is peer-to-peer electricity trading empowering users?: Evidence on motivations and roles in a prosumer business model trial in Australia. Energy Res. Soc. Sci. 66, 101500. 10.1016/j.erss.2020.101500.

Wittenberg, I., Matthies, E., 2016. Solar policy and practice in Germany: How do residential households with solar panels use electricity? Energy Res. Soc. Sci. 21, 199–211. 10.1016/j.erss.2016.07.008.

# Appendix A. Questionnaire

Table A.1 Questions to identify prosumer characteristics.

| | |
|---|---|
| **1. Do you have a renewable energy generating system installed at your home (or another building that belongs to you), such as rooftop PV panels?** | |
| 1.1 | Yes |
| 1.2 | No |
| **2. What was/were the reason(s) you installed that renewable system?** **(multiple answers possible)** | |
| 2.1 | Tackling the climate change problem by being part of the clean energy transition |
| 2.2 | Having more control over my own energy production and use |
| 2.3 | Reduce energy costs |
| 2.4 | There was/is a subsidy |
| 2.5 | Interest in the underlying technologies |
| 2.6 | Other: |
| **3. Are you aware about how much surplus energy you generate with this system?** | |
| 3.1 | Yes |
| 3.2 | No |
| **4. How much surplus energy did you have at the end of last year?** | |
| 4.1 | _______ kWh |
| 4.2 | I don't know |
| **5. Why did you choose your current energy provider?** **(multiple answers possible)** | |
| 5.1 | Green energy supply |
| 5.2 | Lowest cost |
| 5.3 | The company's reputation |
| 5.4 | Bonus for becoming a new customer |
| 5.5 | Provided technology (e.g. an energy app) |
| 5.6 | Recommendation by family or friends |
| 5.7 | Other: |
| **6. Are you a member of an energy cooperative or community** | |
| 6.1 | Yes |
| 6.2 | No |
| **7. Why did you become a member of an energy cooperative or community?** **(multiple answers possible)** | |
| 7.1 | Tackling the climate change problem by being part of the clean energy transition |
| 7.2 | Decentralise energy production |
| 7.3 | Create a local sense of community |
| 7.4 | To have access to renewable energy source technologies |
| 7.5 | Reduce energy costs |
| 7.6 | There is/was a subsidy |
| 7.7 | Improve revenues of our collective or community |
| 7.8 | Other |
| **8. Would you be willing to give surplus energy for free to…** **(multiple answers possible)** | |
| 8.1 | Someone in your neighbourhood/community you don't know |
| 8.2 | Someone in your neighbourhood/community you know |
| 8.3 | A public facility in your community, i.e. school, swimming pool, youth centre |
| 8.4 | A household that can afford energy |
| 8.5 | A friend |
| 8.5 | A family member |
| 8.7 | No, I would not give away my surplus energy for free |

**9. Would you be willing to give surplus energy in exchange for an indirect monetary return (e.g. free entrance to a public facility like museums or swimming pools, a household service like babysitting or gardening)**
**(multiple answers possible)**

| | |
|---|---|
| 9.1 | Someone in your neighbourhood/community you don't know |
| 9.2 | Someone in your neighbourhood/community you know |
| 9.3 | A public facility in your community, i.e. school, swimming pool, youth centre |
| 9.4 | A household that can afford energy |
| 9.5 | A friend |
| 9.6 | A family member |
| 9.7 | No, I would not give away my surplus energy for an indirect monetary return |

Table A.2 Introduction and example of the Discrete Choice Experiment.

In this section, we will be asking you about energy trading. Energy trading happens when you as the person responsible for the system generating electricity with photovoltaic panels or other sources, sell and purchase electricity from and to an electricity grid. Imagine you had to decide between three options regarding the electricity trading in which your household participates. The two trading systems have varying characteristics regarding these six parameters:

- **$CO_2$ emissions:** Electricity can be generated from different sources, which emit different amounts of $CO_2$, which is one of the substances contributing to climate change. Electricity from renewable sources like PV or wind has relatively low $CO_2$ emissions, while electricity generated from coal or gas has relatively high emissions.
- **Electricity tariff**: The electricity tariff determines for how much money you can sell or buy one kilowatt hour of electricity.
- **Social connection with energy trading partner**: Currently, most people who produce their own electricity do not know where their surplus ends up when they feed it into the grid and to their provider. But there can be options where you can decide who you trade your energy with, for example a family member, friend or neighbour.
- **Additional effort:** Choosing an energy provider and selecting a trading system can entail some extra work on a monthly basis. For example, for selecting energy trading partners, observing the ongoing trades or for configuring the technological device.
- **Improved efficiency:** Due to the nature of electricity grids, not all electricity can be utilized and therefore some is always lost. These losses vary depending on the location and management of the grid and can be minimized when additional measures are taken in an electricity trade.
- **Self-sufficiency:** Depending on the set-up of the electricity grid, you as an electricity producing household are dependent on grid operator and energy providers, who can decide over the nature of the electricity trade and have access to your data. Depending on the trading system, there is the possibility to increase the autonomy from them and to have more data privacy.

Some of the trading systems you will see are not currently available, but we would like you to imagine that they are available today. It is important that you answer in the way you would if you were deciding on a trading system.

Example DCE
1 of 12: Please choose one of the following three options.

| | Option 1 | Option 2 | Option 3 |
|---|---|---|---|
| **$CO_2$ emissions** | Low | High | Medium |
| **Social connection with electricity trading partner** | None (unknown) | Direct (neighbour) | Close (friends and family) |
| **Electricity tariff** | 15 ct/kWh | 10 ct/kWh | 20 ct/kWh |
| **Additional effort** | 4 hours/month | 0 h/month | 2 h/month |
| **Improved efficiency** | 15% | 30% | 5% |
| **Self-sufficiency** | Medium | Low | High |

Table A.3 Statements to identify prosumers' attitudes towards the environment, their local community, and technologies.

| | |
|---|---|
| **Environmental** <br> I am concerned about human behaviour and its impact on the climate and the environment. <br> I always pay attention to ecological criteria when buying products and services. | (1) strongly disagree <br> (2) disagree <br> (3) neutral <br> (4) agree <br> (5) strongly agree |
| **Community** <br> I feel a strong identification with my local community. <br> There are many people in my local community whom I think of as friends. | |
| **Technology** <br> Learning how to use a technological device is easy for me. <br> I enjoy exploring new technologies. | |

Table A.4 Questions to identify prosumers' socio-demographic background.

**1. How old are you?**
1.1 Under 25
1.2 25 to 35
1.3 36 to 45
1.4 46 to 55
1.5 56 to 65
1.6 Over 65

**2. What gender do you identify as?**
2.1 Female
2.2 Male
2.3 Third gender/ Non-binary
2.4 Prefer not to say

**3. What is the highest degree or level of education you have achieved?**
3.1 No formal education
3.2 High school diploma
3.3 Vocational training
3.4 Bachelor's degree
3.5 Master's degree
3.6 PhD or higher

| | | |
|---|---|---|
| 3.7 | Prefer not to say | |
| **4. Where is your household located?** | | |
| 4.1 | Large city | |
| 4.2 | Medium-sized city | |
| 4.3 | Small town | |
| 4.4 | Rural community | |
| **5. How many persons live in your household?** | | |
| 5.1 | 1 | |
| 5.2 | 2 | |
| 5.3 | 3 | |
| 5.4 | 4 | |
| 5.5 | 5 or more | |
| **6. What is your average yearly household net income?** | | |
| 6.1 | Under €20,000 | |
| 6.2 | €20,000 to €39,999 | |
| 6.3 | €40,000 to €59,999 | |
| 6.4 | €60,000 to €79,999 | |
| 6.5 | Above €80,000 | |